\documentclass[11pt]{article}
\usepackage{graphicx}

\newcommand{\expect}[1]{{\langle #1 \rangle}}
\newcommand{\maxCoord}{\expect{q_{max}}}
\newcommand{\nextMaxCoord}{\expect{q_{nextMax}}}
\newcommand{\avgCoord}{\expect{q_{avg}}}	
\newcommand{\bdyCoord}{\expect{q_{bdy}}}	
\newcommand{\numEnd}{\expect{N_{end}}}		

\begin{document}

\begin{flushright}
LAUR-00-3925\vspace*{-1ex}\\
SU-4240-722\vspace*{-1ex}\\
August 2000
\end{flushright}

\begin{center}
{\Large \bf Phase diagram of four-dimensional dynamical triangulations
 with a boundary}
 
{\small Simeon Warner$^*$, Simon Catterall$^{\dagger}$ \\
  $^*$ T-8, Los Alamos National Laboratory, NM 87545, USA\\
  $^{\dagger}$ Department of Physics, Syracuse University, 
               Syracuse, NY 13210, USA\\}
\footnotetext{Corresponding author: Simeon Warner, 
email: {\tt simeon@lanl.gov}}
\end{center}

\begin{abstract}
We report on simulations of DT simplicial gravity for manifolds with the
topology of the 4-disk. We find evidence for four phases in a two-dimensional
parameter space. In two of these the boundary plays no dynamical role and
the geometries are equivalent to those observed earlier for the
sphere $S^4$. In another
phase the boundary is maximal and the quantum geometry degenerates to a one
dimensional branched polymer. In contrast we provide evidence that the fourth
phase is effectively {\it three-dimensional}. We find {\it discontinuous}
phase transitions at all the phase boundaries.
\end{abstract}

\section*{Introduction}  
Dynamical triangulation (DT) models arise from
simplicial discretizations of continuous Riemannian manifolds.
A manifold is approximated by glueing together a set of
equilateral simplices with fixed edge lengths. This glueing
ensures that each face is shared by exactly two distinct simplices --
the resultant simplicial lattice is often called a triangulation.
In the context of Euclidean quantum gravity it is
natural to consider a weighted sum of all possible triangulations as a 
candidate for a regularized path integral over metrics. Physically
distinct metrics correspond to inequivalent simplicial
triangulations. This prescription
has been shown to be very successful in two-dimensions in which analytic
methods have been complemented by simulation studies. 
(see, for example,~\cite{AMBJORN94}).  
In dimensions above two there are no known analytic techniques for handling the
sum over triangulations and we must rely on numerical simulation. All of the
latter studies have focused on elucidating the phase structure of {\it compact}
manifolds, principally the sphere $S^4$. In this paper we investigate
the phase structure of four manifolds with the topology of a $4$-disk - that
is a four sphere $S^4$ equipped with a single boundary with
topology $S^3$. As we will demonstrate the dimension of
the parameter space of this model is larger than the corresponding compact
models. The simplest compact models exhibit only
{\it discontinuous} phase transitions precluding a continuum
limit. One motivation for the current work was to see whether the richer parameter
space of the non-compact models contains any continuous phase transitions.  
A natural lattice action $S_b$ can be derived from the continuum action
by straightforward techniques \cite{HARTLE+81}. It contains
both the usual Regge curvature piece familiar from
compact triangulations together with a boundary term.
The boundary term arises from discretization of the extrinsic
curvature of the boundary embedded in the bulk.
In four-dimensions the curvature is localized on triangles. 
If $T_M$ denotes the set of triangles in the
bulk of the 4-triangulation (excluding the boundary) and $T_{\partial M}$
those in the boundary the action can be written 
\begin{equation}
  S_{EH} = \kappa_2\left(\sum_{h\in T_M}\left(2\pi-\alpha n_h\right)+
                        \sum_{h\in T_{\partial M}}
\left(\pi-\alpha n_h\right)\right)+\kappa_4N_4						
\end{equation} 
The quantity $\alpha=\arccos{(1/4)}$, $N_4$ is the 4-volume
and $n_h$ is the number of simplices
sharing the triangle (hinge) $h$. The curvature part of
the action can be rewritten
in terms of the number of vertices $N_0$, the boundary volume
$N_3^b$ and the number of boundary vertices $N_0^b$. With this
in mind we shall consider the general simplicial action
\begin{equation}
  S_{b} = - \kappa_0 N_0 + \kappa_4 N_4 + \kappa_b N_3^b +\kappa_b^0 N_0^b
\label{4daction}
\end{equation}
This form of the action contains both cosmological constant terms and
curvature terms for
the bulk and boundary. Notice that the term involving $\kappa_b^0$ is
absent in three dimensions since it is related to the curvature of a boundary
two-sphere which is a topological invariant.
In this work we have always set $\kappa_b^0=0$ and $\kappa_4$ is used to
tune the volume of the 4-disk.
We are thus left
with a two-dimensional phase space parameterized by $\kappa_0$ 
and $\kappa_b$ conjugate to the number of vertices and the number of
boundary tetrahedra.

The partition function for the system is then
\begin{equation}
  Z = \sum_{T} e^{-S_{b}}
\end{equation}
where the sum is over triangulations, $T$.

\section*{Simulation}

Our simulation algorithm is an extension of the algorithm for 
compact manifolds in arbitrary dimension described by 
Catterall~\cite{CATTERALL95} and is described in~\cite{WARNER+98b}.
To simulate a triangulation with a boundary we actually simulate
a compact triangulation with the topology of
the sphere but consider one {\it marked vertex} to
lie {\it outside} the bulk triangulation. This vertex may never be removed
during the course of the simulation and the surface of the ball created
by its neighbour simplices constitutes the triangulated boundary of the
$4$-disk. The form of the action (equation~\ref{4daction}) can then be derived
by writing down an expression for the integrated curvature of the full
sphere expressed as contributions from the 4-disk and the simplices around
the marked vertex. The latter contributions can be evaluated explicitly in terms
of bulk and boundary simplex numbers and yield this simple form for
the action. To perform a simulation we need a set of local moves which
are ergodic on the space of triangulations of the 4-disk. Fortunately we
have such a set of moves - the usual moves on the sphere.  

In four-dimensions there are just 5 types of move: vertex insertion,
vertex deletion, exchange of a link with a tetrahedron (two moves: 
link to tetrahedron and tetrahedron to link), and exchange of one 
triangle for another triangle. Where these moves take place on 
sections of the triangulation involving the marked vertex we take 
care to count changes in the numbers of simplices inside and 
outside the boundary so that we can calculate the change in the
action, but otherwise the moves are the same as for the bulk.

The code is described in more detail in~\cite{WARNER+98b} where it was used
for the simulation of three-dimensional dynamical triangulations with
a boundary. The code was written for arbitrary dimension and earlier
checked against other workers' results in two-dimensions
\cite{ADI+93}.

We have used the Metropolis Monte Carlo~\cite{METROPOLIS+53} 
scheme with usual update rule:
\begin{equation}
  p({\rm accept\ move}) = min \{ e^{-\Delta S_{b}} , 1 \}
\end{equation}
and in this way we explore the space of triangulations
with the action $S_{b}$.

\subsection*{Measurements}

Here we define some of the measurements used to characterize the 
configurations obtained from our simulations. During the simulations
we store configurations at some interval which is sufficient to 
ensure that the configurations are independent. We check this
by estimating the auto-correlation time ($\tau$) of each measurement
when calculating expectation values. Uncorrelated data would give
$\tau = 0.5$, we quote $\tau$ when it significantly exceeds $0.5$.
 
We use two geodesic measures, $d_{avg}$ and $d_{bdy}$. 
The average geodesic distance between simplices, $d_{avg}$, 
is measured by counting the smallest number of
steps between adjacent simplices required to get from one randomly 
selected simplex to another randomly selected simplex and taking 
the mean over a sample of such measurements. 
The mean geodesic distance to the boundary, $d_{bdy}$, 
is measured by counting the smallest number of
steps between adjacent simplices required to get from one randomly 
selected simplex to the boundary and taking 
the mean over a sample of such measurements. The last step from a 
boundary simplex to the boundary is counted as $0.5$.

To discuss the singular vertex structure observed in some phases
we will consider a {\it coordination} measure, $q$, defined as the
number of simplices sharing a given vertex. 
Similarly we define 
$\maxCoord$ as the expectation value of the coordination
of the most highly coordinated vertex;
$\nextMaxCoord$ as the expectation value of the coordination
of the next most highly coordinated vertex;
$\avgCoord$ as the expectation value of the mean coordination
of all vertices in the triangulation; and
$\bdyCoord$ as the expectation value of the coordination
of the boundary by which we mean the average number of unique 4-simplices
being shared by a typical boundary vertex.

\section*{Phase diagram}

We performed a set of simulations in four dimensions with 
action of equation~\ref{4daction}. In all runs 
$\kappa_4$ was used to tune the nominal system volume, $N_4$,
for each given $\kappa_0$ and $\kappa_b$.
To map the phase diagram we used $N_4 = 1000$; while to
characterize the phases we used
$N_4 = 8000$. In order to check the orders of transitions we have also
used finite size scaling using additional simulations at $N_4=2000$ and
$N_4=4000$.

\begin{figure}[ht]
\begin{center}
\includegraphics[width=11cm]{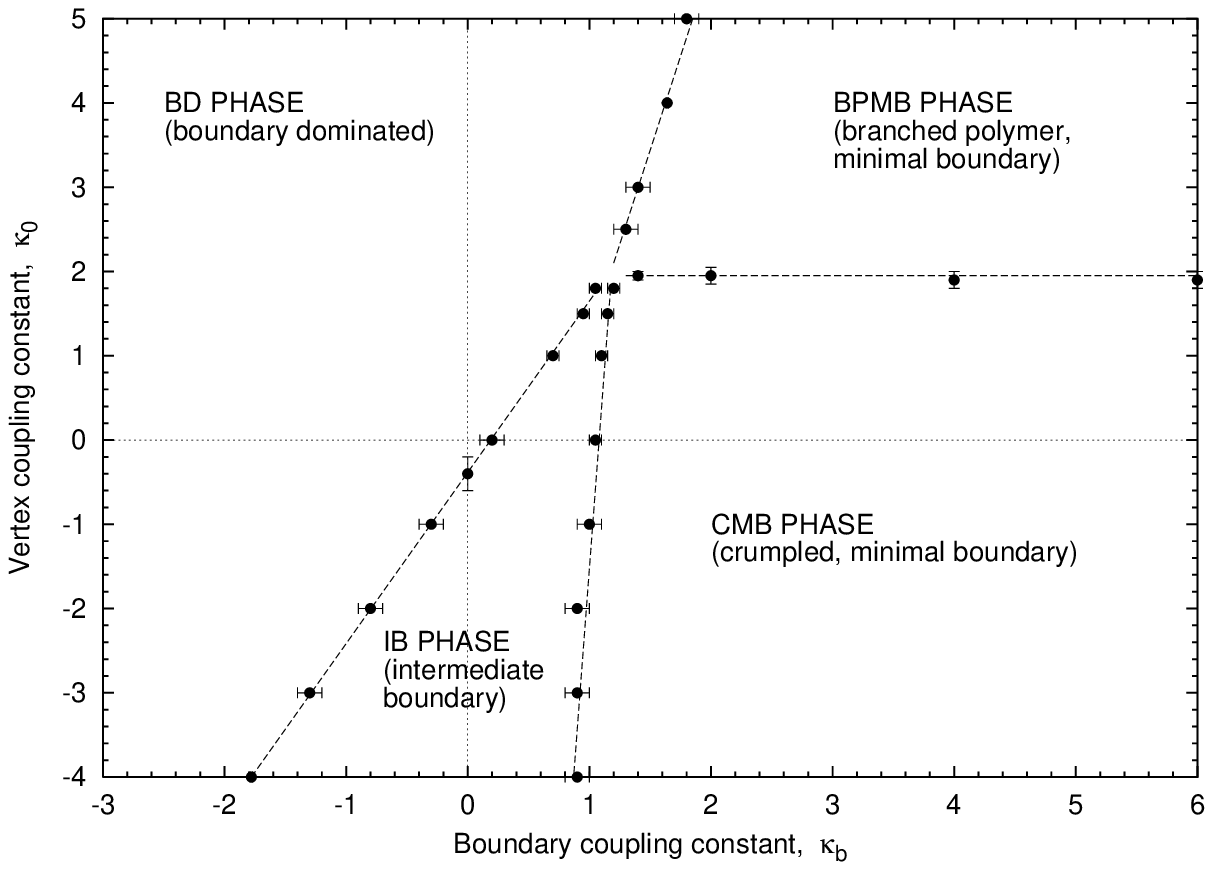}
\end{center}
\caption{\label{phaseplot} Phase diagram for 4-dimensional 
dynamical triangulation with a boundary.
Nominal simulation volume, $N_4 = 1000$. Error bars are
from estimation of the positions of the susceptibility peaks. Where error
bars cannot be seen they are smaller than the symbols; the lines are
guides to the eye.}
\end{figure}

Series of runs varying either $\kappa_0$ or $\kappa_b$ were made
and the vertex susceptibility used to search for phase transitions.
We define the vertex susceptibility, $\chi$, to be normalized with
respect to the number of 4-simplices:
\begin{equation}
  \chi = \frac{1}{N_4} ( \langle N_0^2\rangle - \langle N_0\rangle^2 ) 
\end{equation}
The points shown in figure~\ref{phaseplot} are taken from the positions
of peaks in the vertex and boundary susceptibilities.

\begin{figure}[ht]
\begin{center}
\includegraphics[width=11cm]{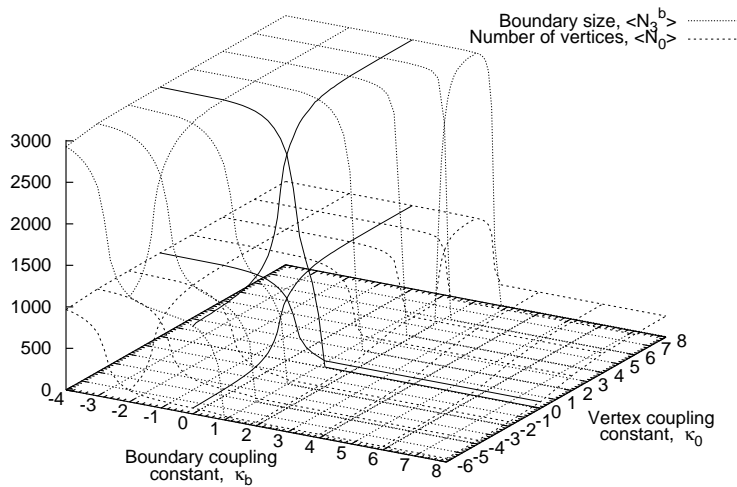}
\end{center}
\caption{\label{3dphaseplot} Boundary size ($\expect{N_3^b}$) and number of
vertices ($\expect{N_0}$) for $4$-dimensional dynamical triangulation 
with a boundary. Nominal simulation volume, $N_4 = 1000$. }
\end{figure} 

In figure~\ref{phaseplot} there are four phases which we characterize as:
crumpled, minimal boundary (CMB); 
branched-polymer, minimal boundary (BPMB);
boundary dominated (BD); and 
intermediate boundary (IB).

In CMB and BPMB phases the boundary is simply 5 tetrahedra (3-simplices) 
connected to form a hyper-tetrahedral hole. The system is essentially a 
sphere with one marked 4-simplex --- the hyper-tetrahedral hole.

\subsection*{CMB phase}

Here we show typical data for $N_4=8000$ at $\kappa_0= -4$ and
$\kappa_b= 4$ using 137 samples (a sample
corresponds to 100000 attempted updates).
The boundary size is just 5 - a minimal hole and we see
one `singular' vertex with $\maxCoord = 2642  (6)$ (the next most coordinated vertex
has $\nextMaxCoord = 531 (5)$, and $\avgCoord = 250.7 (5)$)\\

The presence of singular vertices is similar to the behaviour seen in compact
triangulations in 4-dimensions~\cite{CATTERALL+96, CATTERALL+97}. In the
compact case, there are two singular vertices which are equally coordinated.
In simulations with a boundary we find one singular vertex and the 
minimal-boundary appears to assume the role of the other singular vertex
(the number of simplices shared by boundary vertices is $1871(82)$).

We find $\expect{d_{bdy}} = 9.18 (4)$  (some correlation, $\tau=1.7$) 
and $\expect{d_{avg}} = 10.66 (1)$. The approximate equality of these
two distances tells us that  
in this phase the presence of the boundary plays no crucial role (except
for changing the number of singular vertices). Furthermore, we have
observed that the typical
manifold is very crumpled, having only a logarithmic growth of
its mean size. Such a behavior may be characterized by a large effective
dimension and is reminiscent of the usual crumpled phase seen in
simulations of the sphere.

\subsection*{BPMB phase}

Again we characterize the phase using
data for $N_4=8000$ (116 samples have been acquired)
with couplings $\kappa_0 = 5$, $\kappa_b = 6$.
We find
$\expect{d_{bdy}} = 43.3 (10)$
and $\expect{d_{avg}} = 45.6 (5)$. Again, the distance to the boundary
is comparable to the distance to any randomly selected simplex indicating
that the former has no distinguished role in the triangulation.
Furthermore,
large mean-geodesics are consistent with paths constrained to follow
long branches. 
In this phase the vertex coordinations show no sign of singular structure: 
$\maxCoord = 387 (7)$ and the next most highly coordinated vertex 
is about 50 lower, $\avgCoord = 20.90 (1)$. Thus,
the physics of this phase is again rather like the corresponding 
situation on the sphere, the typical geometries are one-dimensional
polymers.

\subsection*{BD phase}

In this phase the boundary is essentially maximal, the maximal value being 
obtained from an arbitrarily branched chain of simplices such that
$N_{3 maximal}^b = 3\times N_4 + 2$. This is easy to understand in terms of 
a single chain of simplices -- the 2 end 4-simplices have 4 boundary 
3-simplices, and the $(N_4-2)$ middle 4-simplices each have 3 boundary 
3-simplices. 
If one then considers adding a branch, one boundary simplex is
lost at the branch point but one boundary simplex is gained at the 
end simplex, so there is no net change in $N_3^b$.

To support this we list data from a run at $N_4=8000$
and couplings $\kappa_0 = -4$, $\kappa_b = -10$, in which we have
collected 167 samples. We find that the boundary size fluctuates 
only very slightly about a mean of
$N_3^b \approx 24002 = 3 * 8000 + 2$ and is very branched, possessing 
$\numEnd = 2534 (2)$ end points. The maximum vertex coordination
$\maxCoord = 103 (2)$ is close to that expected for a flat lattice.
Another indicator that the geometry corresponds to narrow tubes is seen when
we examine the geodesic distances:
$\expect{d_{bdy}} = 0.5038 (2)$ and
$\expect{d_{avg}} = 125.8 (2)$  
The mean boundary distance tells us that every simplex is a boundary
simplex. Thus the quantum geometry is effectively one-dimensional.
We find $\avgCoord = 4.9975000 (1)$ which we take as further indication 
that there are no sections of bulk in this phase. The argument for this
is as follows: 
in a minimal-width chain with no branches we have two ends which have
a few vertices shared by less than 5-simplices --- the coordination
of vertices in the bulk of the chain. At each end there is 1 vertex
with coordination 1, 1 with coordination 2, 1 with coordination 3,
and 1 with coordination 4. This adds up to a coordination deficit
of 10 for the 4 vertices at an end, 20 for both ends. 
In a minimal-width chain consisting of $N_4$ simplices we expect 
$N_0=N_4+4$ and so with $N_4=8000$  we expect, and see, $N_0=8004$. 
This allows us to calculate the expected 
$\avgCoord = (N_0 \times 5 - 20) / N_0$, putting $N_0=8004$
we get $\avgCoord = 4.997501$. Adding branches to the chain results
in more highly coordinated vertices at the branch point that
exactly cancel the deficit from the additional end.

\subsection*{IB phase}

\begin{figure}[ht]
\begin{center}
\includegraphics[width=9cm]{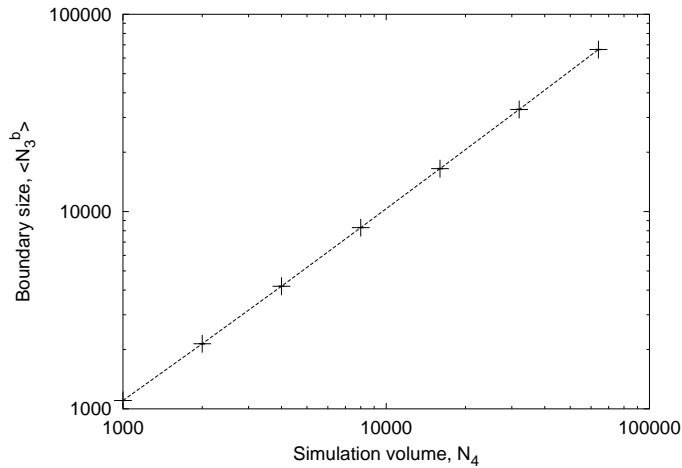}
\end{center}
\caption{\label{scalingplot} Plot showing scaling of boundary size with 
simulation volume for the intermediate boundary (IB) phase, 
$\kappa_0 = -4$, $\kappa_b = 0$. The error bars are much smaller
than the symbols.}
\end{figure}

Here we show data for $N_4=8000$ (86 samples) at couplings
$\kappa_0 = -4$, $\kappa_b = 0$. We find that, once again, the
boundary size scales linearly with the volume (see figure~\ref{scalingplot}) 
yielding $N_3^b \approx N_4\times 1.033$. 
We also find one large
singular vertex, with $\maxCoord = 5721 (9)$ 
($\nextMaxCoord = 233 (2)$, $\avgCoord=133 (1)$).
The measurements of geodesic distances show that the boundary plays 
a preferred role in the triangulation 
$\expect{d_{bdy}} = 0.834 (2)$ while
$\expect{d_{avg}} = 11.36 (1)$. These numbers are incompatible with
an extremal branched polymer shape. However, these measurements by
themselves would not be inconsistent with a `fat-branch' model. If we
add to this the presence of a single singular vertex and the observation that      
about 40\% (3495 (9) out of 8000) of the simplices have 
one boundary face we tentatively conclude that the typical geometry in this
phase is {\it three-dimensional} --- the boundary of the system coinciding with
the boundary of the 4-ball surrounding the singular vertex.


\section*{Phase transitions}

The phase transition between the CMB and BPBM phases has been studied in the
simulations of compact systems in three and four-dimensions and found 
to be discontinuous (first-order) in both cases
(3d~\cite{AMBJORN+92b}, 4d~\cite{BIALAS+96,DEBAKKER96}).
Our simulations have minimal boundary in both phases and so we expect the
same behaviour. This was verified in three dimensions~\cite{WARNER+98b}.

\begin{figure}[ht]
\begin{center}
\includegraphics[width=14cm]{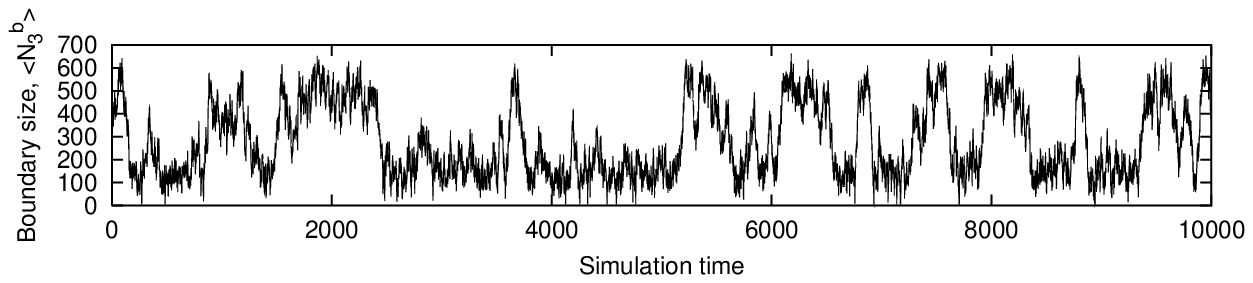}
\includegraphics[width=14cm]{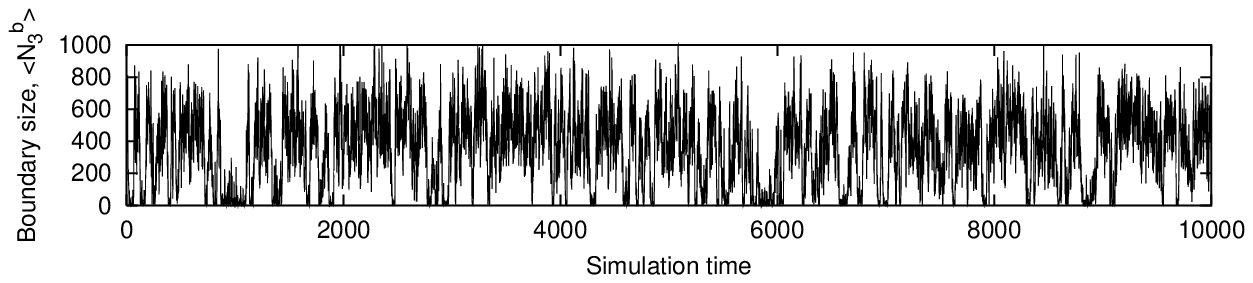}
\includegraphics[width=14cm]{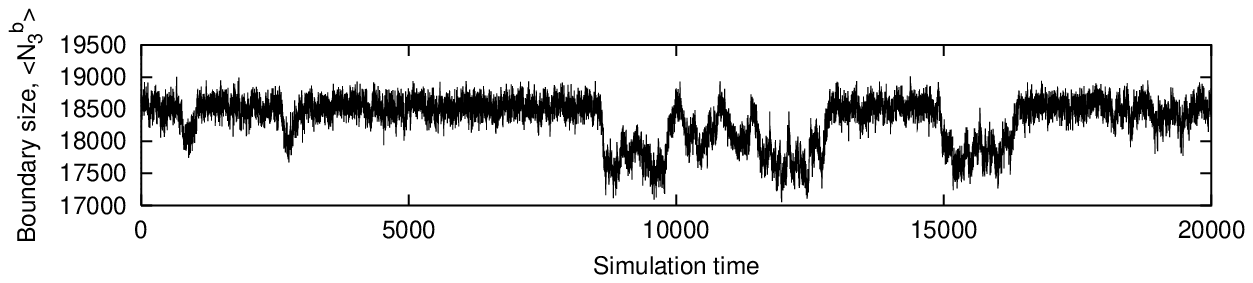}
\end{center}
\caption{\label{bistab-plot} Simulation time series showing bistability 
in the boundary size ($N_3^b$) at 
the CMB---IB phase boundary (upper; $N_4=1000$, $\kappa_0=-4$, $\kappa_b=0.9$); 
the BPMB---BD phase boundary (middle; $N_4=1000$, $\kappa_0=4$, $\kappa_b=1.64$); and 
the BD---IB phase boundary (lower; $N_4=8000$, $\kappa_0=-4$, $\kappa_b=-1.862$). 
We take this as indication of the discontinuous nature of these transitions.}
\end{figure}

We have also investigated the transition between CMB and IB phases and find
good evidence here also for a discontinuous phase transition 
(figure~\ref{bistab-plot}, top; $N_4=1000$, $\kappa_0=-4$, $\kappa_b=0.9$). 
Similarly, the time series close to the BPMB---BD phase boundary
(figure~\ref{bistab-plot}, middle; $N_4=1000$, $\kappa_0=4$, $\kappa_b=1.64$) also
shows signs of bistability indicative of a discontinuous transition.

At $N_4=1000$, 2000 and 4000 we found no such signals near the 
BD---IB boundary. However, we found linear scaling of the
height of the peak in the boundary-size susceptibility which indicates
a first-order transition. To confirm this we extended our simulations 
to $N_4=8000$ and found bistability in the time series 
(figure~\ref{bistab-plot}, bottom; 
$N_4=8000$, $\kappa_0=-4$, $\kappa_b=-1.862$).

\section*{Concluding remarks}

We have simulated four-dimensional simplicial gravity on manifolds with
the topology of a 4-disk. Our action contains both bulk curvature terms and
a boundary cosmological constant term. 
We have identified four phases in the model
within the range of couplings
$-6 < \kappa_0 < 6$ and $-4 < \kappa_b < 8$.
The observed phases include the crumpled and branched-polymer phases
seen in triangulations of compact manifolds, and the boundary
dominated phase seen in three-dimensions. The latter
consists of a maximally branching tree. We have also identified a
fourth phase which 
resembles a singular $4$-ball in which essentially all simplices share
a common bulk vertex and the physics is dominated by the three-dimensional
boundary.

All of the boundaries between these phases appear to be associated with
discontinuous phase transitions. This is the same situation as for 
four-dimensional simplicial gravity on manifolds with $S^4$ topology
and means that we cannot take a continuum limit in the vicinity of
any of the observed phase transitions. Notice that all four phase
boundaries appear to meet (within our errors) at one unique point. The simplest
explanation for this feature is to assume that our phase diagram
contains only two independent transition lines which would then
generically intersect at a single point. Glancing at the nature of the
phases one would associate one line with a boundary dividing lattices
with singular vertices from those without. The second line would then
correspond to a dividing line between geometries with extended 
boundary and those with minimal boundary.
 
Notice also, that in all the phases we have identified, the number
of vertices on the boundary is strongly correlated with the boundary
volume. This need not be the case and it would be interesting to vary the
coupling $\kappa_b^0$ away from zero to see whether a phase could be found which
exhibited a classical scaling of boundary size with 4-volume.

\section*{Acknowledgments}

Simon Catterall was supported in part by DOE grant DE-FG02-85ER40237. 
Simeon Warner was supported in part by the DOE under contract
W-7405-ENG-36.


\bibliographystyle{simeon}
\bibliography{../bib/tcm}

\end{document}